\documentclass[12pt,thmsa]{article}
\usepackage{amsmath}

\usepackage{sw20lart}

\newcommand{\Real}{\operatorname{Re\,}}
\newcommand{\Imag}{\operatorname{Im\,}}

\begin{document}

\author{Emilio Santos \and Departamento de F\'{i}sica. Universidad de Cantabria.
Santander. Spain}
\title{On the analogy between stochastic electrodynamics and nonrelativistic
quantum electrodynamics}
\date{October, 19, 2022 }
\maketitle
\tableofcontents

\begin{abstract}
I expose nonrelativistic quantum electrodynamics in the Weyl-Wigner
representation. Hence I prove that an approximation to first order in Planck
constant has formal analogy with stochastic electrodynamics (SED), that is
classical electrodynamics of charged particles immersed in a random
radiation filling space. The analogy elucidates why SED agrees with quantum
theory for particle Hamiltonians quadratic in coordinates and momenta, but
fails otherwise.
\end{abstract}

\section{Stochastic electrodynamics}

Stochastic electrodynamics (SED in the following) is a theory that studies,
within classical electrodynamics, the motion of charged particles
interacting with electromagnetic fields. The difference with the standard
classical theory is the assumption that there is a random electromagnetic
radiation filling space. This radiation is treated as a stochastic field
whose statistical properties are homogeneous, isotropic and Lorentz
invariant. SED may be labelled a classical (i. e. not quantum) theory,
although the assumption of a random background radiation is alien to, but
complatible with, classical electrodynamics.

A reasoning leading to SED is as follows. When Rutherford proposed the
nuclear atom, a common argument against the model was that such atom could
not be stable. In fact the accelerated electron should radiate whence the
atom would collapse. But this argument is misleading\cite{Santos68}, the
atom could not be stable \textit{if isolated}. However the hypothesis of
isolation is not appropriate if there are many atoms in the universe. It is
more plausible to assume that there is some amount of radiation filling
space, whence every atom would radiate but it could also absorb energy from
the radiation, eventually arriving at a dynamical equilibrium. This might
explain the stability of the atom. Of course Bohr had solved the problem
introducing postulates actually incompatible with classical laws. Early
workers believed that SED, after some elaboration, might give rise to a
reinterpretation of quantum mechanics. However the attempt has not been
successful.

In order to study the assumed radiation of SED we may expand it in plane
waves, within a large volume, and represent the amplitudes of the waves by
dimensionless complex numbers $\left\{ a_{l}\right\} $. It is plausible that
different amplitudes are statistically uncorrelated and the probability of
every one is Gaussian. Hence the probability distribution of the radiation
amplitudes is as follows 
\begin{equation}
W_{vac}=\prod_{l=1}^{n}\frac{2}{\pi }\exp \left( -2\left| a_{l}\right|
^{2}\right) ,  \label{Wvac}
\end{equation}
normalized with respect to the integration in $d\Real a_{l}$ $d\Imag %
a_{l}.$

In free space it is plausible that the spectral energy density, $\rho
_{SED}\left( \omega \right) ,$ (energy per unit volume and unit frequency
interval, in short ``spectrum'') of the random radiation is homogeneous,
isotropic and Lorentz invariant in the following sense. In a given inertial
frame the frequency of one plane wave may change when we observe the
radiation in a different inercial frame so that, fixing a small frequency
interval $\left( \omega _{1},\omega _{2}\right) ,$ some plane wave
frequencies will enter the interval and other frequencies will leave it with
the change of frame. We define the spectrum as Lorentz invariant if the
number of frequencies that enter the interval $\left( \omega _{1},\omega
_{2}\right) $ balances the number that leave it, this for all frequency
intervals. A necessary and sufficient condition for Lorentz invariance, in
the said sense, is that the spectrum is proportional to the cube of the
frequency \cite{Milonni}, \cite{dice}. Thus the spectrum of the random
radiation may be written 
\begin{equation}
\rho _{SED}\left( \omega \right) =C\omega ^{3},C=\frac{%
%TCIMACRO{\UNICODE[m]{0x127}}
%BeginExpansion
\rlap{\protect\rule[1.1ex]{.325em}{.1ex}}h%
%EndExpansion
}{2\pi ^{2}c^{3}},  \label{roZPF}
\end{equation}
whence Planck constant $%TCIMACRO{\UNICODE[m]{0x127}}
%BeginExpansion
\rlap{\protect\rule[1.1ex]{.325em}{.1ex}}h%
%EndExpansion
$ enters the theory via fixing the scale of the assumed universal radiation.
The spectrum eq.$\left( \ref{roZPF}\right) $ corresponds to an average
energy 
\begin{equation}
\left\langle E_{j}\right\rangle =\frac{1}{2}%TCIMACRO{\UNICODE[m]{0x127}}
%BeginExpansion
\rlap{\protect\rule[1.1ex]{.325em}{.1ex}}h%
%EndExpansion
\omega _{j},  \label{EZPF}
\end{equation}
for the $j$ plane wave in the expansion. Then we may write the total energy, 
$E_{SED}\left\{ a_{j}\right\} ,$ of the SED random field in terms of the
amplitudes demanding that the average of $E_{SED},$ taking eq.$\left( \ref
{Wvac}\right) $ into account, gives eq.$\left( \ref{EZPF}\right) .$ This
leads to the following 
\begin{equation}
E_{SED}=\sum_{l}E_{j}=\sum_{l}%TCIMACRO{\UNICODE[m]{0x127}}
%BeginExpansion
\rlap{\protect\rule[1.1ex]{.325em}{.1ex}}h%
%EndExpansion
\omega _{l}\left| a_{l}\right| ^{2}.  \label{ESED}
\end{equation}
The spectrum eq.$\left( \ref{roZPF}\right) $ implies a divergent total
energy density and any cutoff would break Lorentz invariance, but eq.$\left( 
\ref{roZPF}\right) $ is assumed to be valid for low enough frequencies and
the effect of high frequencies is neglected (it would produce on any charged
particle a rapid shaking superposed to a more smooth path). Ultraviolet
divergences are well known to exist also in QED and they are associated to
the quantum vacuum fluctuations \cite{Milonni}.

A review of the work on SED made until 1995 is the book by L. de la Pe\~{n}a
and A. M. Cetto \cite{dice} and new results are included in more recent
reviews \cite{dice2}, \cite{Boyer19}, \cite{book}, \cite{arxiv}.

In practice SED has been studied within the nonrelativistic approximation
ignoring spin. In order to find the evolution equations of the particles we
shall solve the dynamical (Newtonian) equation of motion that takes the
radiation into account. I will write it in one dimension but might be
extended to 3N dimensions for N-particle systems. That is 
\begin{equation}
m\stackrel{..}{x}=-\frac{dV\left( x\right) }{dx}+m\tau \stackrel{...}{x}%
+eE\left( t\right) ,  \label{ode}
\end{equation}
where $m(e)$ is the particle mass (charge) and $E$ is the $x$ component of
the electric field of the radiation that acts on the particle. The field $%
E\left( t\right) $ is treated as a stochastic process whence eq.$\left( \ref
{ode}\right) $ is a stochastic differential equation of Langevin type with
non-white noise. We see that Newton equation is modified by two terms which
correspond to the matter-radiation interaction. The second term on the right
side is the radiation reaction force due to emission from the charged
paticle. This term does not contain $%TCIMACRO{\UNICODE[m]{0x127}}
%BeginExpansion
\rlap{\protect\rule[1.1ex]{.325em}{.1ex}}h%
%EndExpansion
$ and it is well known in standard classical electrodynamics. The third term
is the (Lorentz) force of the radiation on the particle in the electric
dipole approximation, which is consistent with the nonrelativistic
treatment. The approximation consists of neglecting the action of the
magnetic field and the dependence on position of the electric field. This
term is specific of SED and it is proportional to the square root of Planck
constant because the average of $E(t)^{2}$ is proportional to the spectrum
eq.$(\ref{roZPF}).$ The parameter $\tau $ is given by 
\begin{equation}
\tau =\frac{2e^{2}}{3mc^{3}}\Rightarrow \tau \omega _{0}=\frac{2}{3}\frac{%
e^{2}}{%TCIMACRO{\UNICODE[m]{0x127}}
%BeginExpansion
\rlap{\protect\rule[1.1ex]{.325em}{.1ex}}h%
%EndExpansion
c}\frac{%TCIMACRO{\UNICODE[m]{0x127}}
%BeginExpansion
\rlap{\protect\rule[1.1ex]{.325em}{.1ex}}h%
%EndExpansion
\omega _{0}}{mc^{2}}<<1.  \label{gamma}
\end{equation}
so that the dimensionless quantity $\tau \omega _{0}$ is very small, it
being the product of two small numbers namely the fine structure constant, $%
\alpha \equiv e^{2}/%TCIMACRO{\UNICODE[m]{0x127}}
%BeginExpansion
\rlap{\protect\rule[1.1ex]{.325em}{.1ex}}h%
%EndExpansion
c\sim 1/137,$ and the nonrelativistic ratio $%TCIMACRO{\UNICODE[m]{0x127}}
%BeginExpansion
\rlap{\protect\rule[1.1ex]{.325em}{.1ex}}h%
%EndExpansion
\omega _{0}/mc^{2}\simeq $ $v^{2}/c^{2}<<1.$ It may be shown that the latter
term of eq.$\left( \ref{ode}\right) $ is also small, which simplifies the
solution, but I shall not study further this approach which may be seen in
the references quoted above.

The results obtained from eq.$\left( \ref{ode}\right) $ agree, in the limit $%
\alpha \rightarrow 0,$ with the quantum predictions for quadratic
Hamiltonians. For instance the squared mean position and momentum or the
energy in the ground state are fairly well predicted. Corrections of order $%
\alpha $ may be obtained from eq.$\left( \ref{ode}\right) $ which correspond
to radiative corrections of QED (e.g. Lamb shift). The solution of eq.$%
\left( \ref{ode}\right) $ is easy when it is linear, i.e. for quadratic
Hamiltonians. It is not so easy for nonlinear systems, although many papers
have been devoted to approximate solutions in that case. However the problem
with nonlinear systems is not just the difficulty of the solution, but the
disagreement of the results with the quantum prediction as commented on
below. Explaning this feature is one of the purposes of the present article.

The interest on SED is due to the fact that it provides an interpretation of
several typically quantum phenomena in the spirit of classical physics. For
instance the mentioned stability of the classical (Rutherford) atom,
Heisenberg uncertainty relations, and some examples of entanglement. SED
provides too a classical-like interpretation of phenomena where the spectrum
of free space is modified by boundary conditions derived from macroscopic
bodies, but the average energy $\frac{1}{2}%TCIMACRO{\UNICODE[m]{0x127}}
%BeginExpansion
\rlap{\protect\rule[1.1ex]{.325em}{.1ex}}h%
%EndExpansion
\omega $ per normal mode still holds, e.g. the Casimir effect and the
behaviour of atoms in cavities. SED may be also studied at a finite
temperature, where the thermal Planck spectrum is added to eq.$\left( \ref
{roZPF}\right) $, thus correctly predicting the specific heats of solids.

As the\ study of some simple systems provides an intuitive picture of
several quantum phenomena SED has been proposed as a clue in the search for
a realistic interpretation of quantum theory \cite{arxiv}. A different
approach is to consider that SED is the closest classical approximation to
quantum theory \cite{Boyer19}. Furthermore SED, with appropriate
generalizations, has been believed a possible alternative to quantum
mechanics\cite{AnaLuis}. However there are many examples where SED predicts
results in contradiction with quantum mechanics and experiments. Another
shortcoming is that SED deals only with charged particles whilst QM laws are
valid also for neutral particles.

In this article I will show that SED might be taken as a semiclassical
theory. In fact it has similarity with nonrelativistic QED approximated to
first order in Planck\'{}s constant. This would elucidate why SED gives
predictions in agreement with quantum mechanics (and experiments) for linear
systems (i.e. Hamiltonians quadratic in positions and momenta like the
harmonic oscillator), but not for nonlinear systems.

In order to prove the assertion I will start revisiting the standard Hilbert
space formalism of QED. (Throughout this article I will use QED for \textit{%
nonrelativistic} quantum electrodynamics, the relatistic theory will not be
studied here). After that I will briefly review the Weyl-Wigner formalism
and derive an equation similar to eq.$\left( \ref{ode}\right) $ of SED as an
approximation of order O$\left( %TCIMACRO{\UNICODE[m]{0x127}}
%BeginExpansion
\rlap{\protect\rule[1.1ex]{.325em}{.1ex}}h%
%EndExpansion
\right) $, zeroth order corresponding to classical electrodynamics.

\section{Nonrelativistic quantum electrodynamics}

\subsection{The standard Hilbert space formulation}

For later convenience I will revisit the well known Hilbert space
formulation of nonrelativistic quantum mechanics coupled to the quantized
electromagnetic field.

The fundamental equation should provide the evolution of the state of a
system of particles, those charged interacting with each other, eventually
with external fields, and with the quantized electromagnetic field. The
state is represented by a density operator (or density matrix) $\hat{\rho}$
whose evolution is given by the equation 
\begin{equation}
i%TCIMACRO{\UNICODE[m]{0x127}}
%BeginExpansion
\rlap{\protect\rule[1.1ex]{.325em}{.1ex}}h%
%EndExpansion
\frac{d}{dt}\hat{\rho}=\left[ \hat{H},\hat{\rho}\right] \equiv \hat{H}\hat{%
\rho}-\hat{\rho}\hat{H},  \label{s2}
\end{equation}
where the Hamiltonian consists of 3 terms, that is 
\begin{equation}
\hat{H}=\hat{H}_{part}+\hat{H}_{rad}+\hat{H}_{int},  \label{s3}
\end{equation}
for the particles, the radiation field and the interaction, respectively.
For the sake of clarity I shall represent the operators with a `hat', e.g. $%
\stackrel{\wedge }{x}_{j},\stackrel{\wedge }{p}_{j},$ and the classical
quantities (c-numbers) without `hat', e.g. $x_{j},p_{j}.$

A typical, but not general, particle Hamiltonian is the following

\begin{equation}
\hat{H}_{part}=\sum_{j=1}^{N}\left[ \frac{\mathbf{\hat{p}}_{j}^{2}}{2m_{j}}%
+V\left( \left\{ \mathbf{\hat{x}}_{j}\right\} \right) \right] ,  \label{c1}
\end{equation}
where $m_{j},$ $\mathbf{\hat{x}}_{j}$ and $\mathbf{\hat{p}}_{j}$ are the
position and momentum of the particle $j$, and the potential $V$ should
include the interaction with given external fields and the instantaneous
electrostatic interactions amongst the charged particles, which after the
coupling with the vector potential of the radiation field would provide all
electromagnetic interactions in the Coulomb gauge. Throughout the paper I
will use that gauge, rather than Lorentz's, as appropriate for a
non-relativistic theory.

For the study of radiation a popular approach is to start with a plane waves
expansion of the vector potential operator in a large volume $V$. Hence the
Hamiltonian, representing the radiation energy in that volume$,$ becomes 
\begin{equation}
\hat{H}_{rad}=\frac{1}{2}\sum_{l}%TCIMACRO{\UNICODE[m]{0x127}}
%BeginExpansion
\rlap{\protect\rule[1.1ex]{.325em}{.1ex}}h%
%EndExpansion
\omega _{l}(\hat{a}_{l}\hat{a}_{l}^{\dagger }+\hat{a}_{l}^{\dagger }\hat{a}%
_{l}),  \label{s4}
\end{equation}
where the operators fulfil the commutation rules 
\[
\left[ \hat{a}_{k},\hat{a}_{l}\right] =\left[ \hat{a}_{k}^{\dagger },\hat{a}%
_{l}^{\dagger }\right] =0,\left[ \hat{a}_{k},\hat{a}_{l}^{\dagger }\right]
=\delta _{kl}, 
\]
and the ground state of the field is given by the vector state $\mid
0\rangle $ fulfilling 
\begin{equation}
\hat{a}_{j}\mid 0\rangle =\langle 0\mid \hat{a}_{j}^{\dagger }=0,  \label{h4}
\end{equation}
where $0$ means the nul vector in the Hilbert space. This defines the
minimal energy state of the field.

The interaction Hamiltonian may be written in terms of the vector potential,
that is 
\begin{equation}
\hat{H}_{int}=\sum_{j}[-\frac{e}{mc}\mathbf{\hat{p}}_{i}\cdot \mathbf{\hat{A}%
}\left( \mathbf{\hat{x}}_{j},t\right) +\frac{e^{2}}{2mc^{2}}\mathbf{\hat{A}}%
\left( \mathbf{\hat{x}}_{j},t\right) \cdot \mathbf{\hat{A}}\left( \mathbf{%
\hat{x}}_{j},t\right) ],  \label{Hint}
\end{equation}
where the sum runs over the different particles, $\mathbf{\hat{x}}_{j}$ and $%
\mathbf{\hat{p}}_{j}$ being the position and momentum of particle \textit{j}
at time t. For simplicity I will assume that all particles have the same
mass $m$ and charge $e$ from now on. The formalism sketched allowed
nonrelativistic QED computations able to reproduce empirical results like
the Lamb shift (e.g. the celebrated calculation by Bethe in 1947) \cite
{Milonni}. Of course the theory was superseded with the advent of
relativistic QED.

\subsection{Weyl-Wigner formulation}

The Hilbert space formulation of quantum theory is not convenient for our
purpose of studying the analogy with SED, a better procedure being to work
in the Weyl-Wigner formalism (WW in the following). I stress that both are
different forms of quantum theory that should predict the same results for
experiments. In 1927 Hermann Weyl \cite{Weyl} proposed a general procedure
for the quantization of (nonrelativistic) classical mechanics of particles
via a transform that provides operators corresponding to positions and
momenta of particles. Here we are interested in the inverse Weyl transform
that may be applied to a nonrelativistic system of particles as follows. For
any trace-class operator $\hat{f}$ in the quantum Hilbert space of the
particles the transform gives a function $W$ in the classical phase space as
follows \cite{Scully}, \cite{Zachos} 
\begin{eqnarray}
W\left( \left\{ x_{j}\mathbf{,}p_{j}\right\} \right) &=&\left( \pi \right)
^{-6N}\int Tr\left\{ \stackrel{\wedge }{f}\exp \left[ i\sum_{j=1}^{3N}\left(
\lambda _{j}\stackrel{\wedge }{x}_{j}+\mu _{j}\stackrel{\wedge }{p}%
_{j}\right) \right] \right\}  \nonumber \\
&&\times \exp \left[ (-i\sum_{j=1}^{3N}\left( \lambda _{j}x_{j}+\mu
_{j}p_{j}\right) \right] \prod_{j=1}^{3N}d\lambda _{j}d\mu _{j}.  \label{W}
\end{eqnarray}
In the particular case when $\hat{f}$ is the density operator of a quantum
state the transform eq.$\left( \ref{W}\right) $ gives the Wigner function of
the state \cite{Wigner}. Another application of eq.$\left( \ref{W}\right) $
is to derive observables from the corresponding operators in the Hilbert
space, in particular the particle Hamiltonian, which for eq.$\left( \ref{c1}%
\right) $ gives simply 
\begin{equation}
H_{part}=\sum_{j=1}^{N}\left[ \frac{\mathbf{p}_{j}^{2}}{2m}+V\left( \left\{ 
\mathbf{x}_{j}\right\} \right) \right] .  \label{c2}
\end{equation}
However if $\hat{H}_{part}$ contained products of $\mathbf{\hat{x}}_{i}$
times $\mathbf{\hat{p}}_{i}$ the result of the Weyl transform would not be
so simple due to the non-commutativity of these operators.

The WW formalism also allows getting the evolution of the Wigner function of
a system of particles. For the Hamiltonian eq.$\left( \ref{c2}\right) $ it
is the following (for a single particle, but the generalization to $N$
particles is straightforward) 
\begin{eqnarray}
\frac{\partial W\left( \mathbf{x,p}\right) }{\partial t} &=&-\frac{1}{m}%
\mathbf{p}\cdot \mathbf{\nabla }W-\frac{1}{%TCIMACRO{\UNICODE[m]{0x127}}
%BeginExpansion
\rlap{\protect\rule[1.1ex]{.325em}{.1ex}}h%
%EndExpansion
}\int \frac{d\mathbf{p}^{\prime }}{(2\pi )^{3}}\widetilde{V}\left( \mathbf{%
x,p}^{\prime }\right) W\left( \mathbf{x,p+p}^{\prime }\mathbf{,}t\right) 
\nonumber \\
\widetilde{V}\left( \mathbf{x,p}^{\prime }\right) &\equiv &\int d\mathbf{u}%
\sin \left( \mathbf{p}^{\prime }\mathbf{\cdot u}\right) \left[ V\left( 
\mathbf{x+}%TCIMACRO{\UNICODE[m]{0x127}}
%BeginExpansion
\rlap{\protect\rule[1.1ex]{.325em}{.1ex}}h%
%EndExpansion
\mathbf{u/}2\right) -V\left( \mathbf{x-}%TCIMACRO{\UNICODE[m]{0x127}}
%BeginExpansion
\rlap{\protect\rule[1.1ex]{.325em}{.1ex}}h%
%EndExpansion
\mathbf{u/}2\right) \right] .  \label{4.50}
\end{eqnarray}
In summary the WW formalism allows to study nonrelativistic quantum
mechanics providing indentical predictions than the standard Hilbert space,
that is both formalisms are physically equivalent. However the Wigner
function not always can be interpreted as a probability distribution in
phase space because it is not positive in general.

For our purpose another form of the evolution is more convenient than eq.$%
\left( \ref{4.50}\right) $ which is valid for any particle Hamiltonian, not
just eq.$\left( \ref{c2}\right) $. It is given by the Moyal equation for any
Wigner function $W\left( \left\{ x_{j},p_{j}\right\} \right) $, that is 
\begin{eqnarray}
\frac{\partial W}{\partial t} &=&\frac{2}{%TCIMACRO{\UNICODE[m]{0x127}}
%BeginExpansion
\rlap{\protect\rule[1.1ex]{.325em}{.1ex}}h%
%EndExpansion
}\sin \left[ \frac{%TCIMACRO{\UNICODE[m]{0x127}}
%BeginExpansion
\rlap{\protect\rule[1.1ex]{.325em}{.1ex}}h%
%EndExpansion
}{2}\left( \frac{\partial }{\partial x_{j}}\frac{\partial }{\partial
p_{j}^{\prime }}-\frac{\partial }{\partial p_{j}}\frac{\partial }{\partial
x_{j}^{\prime }}\right) \right] \left[ W\left( \left\{ x_{j}\mathbf{,}%
p_{j}\right\} \right) H_{part}\left( \left\{ x_{j}^{\prime }\mathbf{,}%
p_{j}^{\prime }\right\} \right) \right]  \nonumber \\
{} &\equiv &\frac{2}{%TCIMACRO{\UNICODE[m]{0x127}}
%BeginExpansion
\rlap{\protect\rule[1.1ex]{.325em}{.1ex}}h%
%EndExpansion
}\sum_{n=0}^{3N}\frac{\left( -1\right) ^{n}}{\left( 2n+1\right) !}\left[ 
\frac{%TCIMACRO{\UNICODE[m]{0x127}}
%BeginExpansion
\rlap{\protect\rule[1.1ex]{.325em}{.1ex}}h%
%EndExpansion
}{2}\left( \frac{\partial }{\partial x_{j}}\frac{\partial }{\partial
p_{j}^{\prime }}-\frac{\partial }{\partial p_{j}}\frac{\partial }{\partial
x_{j}^{\prime }}\right) \right] ^{2n+1}  \nonumber \\
&&\times \left[ W\left( \left\{ x_{j}\mathbf{,}p_{j}\right\} \right)
H_{part}\left( \left\{ x_{j}^{\prime }\mathbf{,}p_{j}^{\prime }\right\}
\right) \right] \equiv \left\{ W,H_{part}\right\} _{M},  \label{dW}
\end{eqnarray}
where we should identify $\left\{ x_{j}^{\prime },p_{j}^{\prime }\right\}
=\left\{ x_{j},p_{j}\right\} $ after performing the derivatives. $\left\{
W,H_{part}\right\} _{M}$ is a simplified notation to be used in the
following, the subindex $M$ standing for Moyal bracket. For the Hamiltonian
eq.$\left( \ref{c2}\right) $ the Moyal equation may be derived from eq.$%
\left( \ref{4.50}\right) $ via an expansion in powers of $
%TCIMACRO{\UNICODE[m]{0x127}}
%BeginExpansion
\rlap{\protect\rule[1.1ex]{.325em}{.1ex}}h%
%EndExpansion
$.

Eq.$\left( \ref{dW}\right) $ is valid for any (finite) set of particles with
a general Hamiltonian written in terms of generalized coordinates and
momenta, but it involves derivatives of infinite order, which makes it just
a formal (not practical) equation. However it may provide usefull
aproximations if truncated at a finite order in Planck constant $
%TCIMACRO{\UNICODE[m]{0x127}}
%BeginExpansion
\rlap{\protect\rule[1.1ex]{.325em}{.1ex}}h%
%EndExpansion
$. In the limit $%TCIMACRO{\UNICODE[m]{0x127}}
%BeginExpansion
\rlap{\protect\rule[1.1ex]{.325em}{.1ex}}h%
%EndExpansion
\rightarrow 0$ eq.$\left( \ref{dW}\right) $ reduces to the classical
Liouville equation, Moyal bracket becoming the Poisson bracket.

Weyl transform eq.$\left( \ref{W}\right) $ may be extended to the radiation
field \cite{Frontiers}, \cite{FOOP}, taking into account that some linear
combinations of the ``creation and annihilation operators of photons'' in
the normal modes (i.e. plane waves in free space) may play the role of
``coordinates and momenta'' operators. We may write the following (inverse)
Weyl transform for the radiation field 
\begin{eqnarray}
f\left( \left\{ a_{l}\right\} \right) &=&\left( \frac{2}{\pi }\right)
^{2n}\prod_{l=1}^{n}\int_{-\infty }^{\infty }d\lambda _{l}\int_{-\infty
}^{\infty }d\mu _{l}\exp \left[ -2i(\lambda _{l}\Real a_{l}+\mu _{l}%
\Imag a_{l})\right]  \nonumber \\
&&\times Tr\left\{ \hat{f}\exp \left[ i\lambda _{l}\left( \hat{a}_{l}+\hat{a}%
_{l}^{\dagger }\right) +\mu _{l}\left( \hat{a}_{l}-\hat{a}_{l}^{\dagger
}\right) \right] \right\} ,  \label{h2}
\end{eqnarray}
where $a_{l}$ and $a_{l}^{*}$ are c-number field amplitudes that in the WW
formalism are substituted for the Hilbert space operators $\hat{a}_{l}$ and $%
\hat{a}_{l}^{\dagger }$.

The transform eq.$\left( \ref{h2}\right) $ allows getting the quantum states
and observables in the standard Hilbert space formalism as functions of the
amplitudes $\left\{ a_{l}\right\} .$ Then the quantum field looks like a
classical field in the WW formalism. In particular we may describe the field
by a vector potential, whence we might derive the electric and magnetic
fields and hence the Hamiltonian, that is 
\begin{equation}
H_{rad}=\sum_{l}%TCIMACRO{\UNICODE[m]{0x127}}
%BeginExpansion
\rlap{\protect\rule[1.1ex]{.325em}{.1ex}}h%
%EndExpansion
\omega _{l}\left| a_{l}\right| ^{2}=\sum_{l}%TCIMACRO{\UNICODE[m]{0x127}}
%BeginExpansion
\rlap{\protect\rule[1.1ex]{.325em}{.1ex}}h%
%EndExpansion
\omega _{l}\left[ (\Real a_{l})^{2}+(\Imag a_{l})^{2}\right] ,
\label{rad}
\end{equation}
to be compared with eq.$\left( \ref{s4}\right) .$ As is well known the
evolution of $\Real a_{l}$ and $\Imag a_{l}$ is formally similar to
the position and momentun of a harmonic oscillator, which may be stressed
with the following change of variables which I will use from now on 
\begin{equation}
y_{l}=\sqrt{\frac{2%TCIMACRO{\UNICODE[m]{0x127}}
%BeginExpansion
\rlap{\protect\rule[1.1ex]{.325em}{.1ex}}h%
%EndExpansion
}{\omega _{l}}}\Real a_{l},q_{l}=\sqrt{2%TCIMACRO{\UNICODE[m]{0x127}}
%BeginExpansion
\rlap{\protect\rule[1.1ex]{.325em}{.1ex}}h%
%EndExpansion
\omega _{l}}\Imag a_{l}\Rightarrow H_{rad}=\sum_{l}\left[ \frac{1}{2}%
\omega _{l}^{2}y_{l}^{2}+\frac{1}{2}q_{l}^{2}\right] .  \label{yq}
\end{equation}
Hence the Hamilton equations give, for free radiation, 
\begin{equation}
\frac{d}{dt}y_{l}=\frac{\partial H_{rad}}{\partial q_{l}}=q_{l},\frac{d}{dt}%
q_{l}=-\frac{\partial H_{rad}}{\partial y_{l}}=-\omega _{l}^{2}y_{l}.
\label{s5}
\end{equation}
An advantage of the new variables is that the Hamiltonian $H_{rad},$ eq.$%
\left( \ref{yq}\right) ,$ does not contain Planck constant $
%TCIMACRO{\UNICODE[m]{0x127}}
%BeginExpansion
\rlap{\protect\rule[1.1ex]{.325em}{.1ex}}h%
%EndExpansion
$, at a difference with eq.$\left( \ref{rad}\right) .$ The presence of $
%TCIMACRO{\UNICODE[m]{0x127}}
%BeginExpansion
\rlap{\protect\rule[1.1ex]{.325em}{.1ex}}h%
%EndExpansion
$ might mislead us to believe that eq.$\left( \ref{rad}\right) $ is a
specifically quantum relation. Actually the introduction of $
%TCIMACRO{\UNICODE[m]{0x127}}
%BeginExpansion
\rlap{\protect\rule[1.1ex]{.325em}{.1ex}}h%
%EndExpansion
$ in eq.$\left( \ref{s4}\right) $ is currently made in order that the
quantum operators $\left\{ \hat{a}_{l},\hat{a}_{l}^{\dagger }\right\} $ are
dimensionless but it has no deep meaning. However the choice eq.$\left( \ref
{rad}\right) $ is inconvenient in this paper where we want to distinguish
quantum from classical features.

\subsection{The quantum vacuum or zeropoint field}

In the standard, Hilbert space, formalism the vacuum corresponds to the
absence of ``photons''. The vacuum state $W_{vac}\left( \left\{
a_{j},a_{j}^{*}\right\} \right) $\ in the WW formalism should be derived
from the vacuum density operator in Hilbert space, that is $\left| 0\rangle
\langle 0\right| ,$ via the Weyl transform eq.$\left( \ref{h2}\right) .$ To
do that I shall start calculating the trace operation putting $\hat{f}%
=\left| 0\rangle \langle 0\right| $ in eq.$\left( \ref{W}\right) ,$ taking
the Campbell-Haussdorf formula into account that is 
\begin{equation}
\exp \left( \hat{A}+\hat{B}\right) =\exp \left( \hat{A}\right) \exp \left( 
\hat{B}\right) \exp \left( -\frac{1}{2}\left[ \hat{A},\hat{B}\right] \right)
,  \label{h5}
\end{equation}
which is valid if the operator $\left[ \hat{A},\hat{B}\right] $ commutes
with $\hat{A}$ and with $\hat{B}$. Hence the trace involved in eq.$\left( 
\ref{W}\right) $ becomes, for a single mode, 
\begin{eqnarray*}
&&Tr\left\{ \left| 0><0\right| \exp \left[ i\lambda \left( \hat{a}+\hat{a}%
^{\dagger }\right) +\mu \left( \hat{a}-\hat{a}^{\dagger }\right) \right]
\right\} \\
&=&\langle 0\mid \exp \left[ (i\lambda -\mu )\hat{a}^{\dagger }\right] \exp
\left[ (i\lambda +\mu )\hat{a}\right] \mid 0\rangle \exp \left[ -\lambda
^{2}/2-\mu ^{2}/2\right] \\
&=&\exp \left[ -\lambda ^{2}/2-\mu ^{2}/2\right] .
\end{eqnarray*}
If this is inserted in eq.$\left( \ref{h2}\right) $ we get 
\begin{eqnarray}
W_{vac} &=&\prod_{l=1}^{n}\frac{4}{\pi ^{2}}\int_{-\infty }^{\infty
}d\lambda _{l}\int_{-\infty }^{\infty }d\mu _{l}  \nonumber \\
&&\times \exp \left[ -2i(\lambda _{l}\Real a_{l}-\mu _{l}\mathrm{Im}%
a_{l})-\frac{1}{2}\left( \lambda _{l}^{2}+\mu _{l}^{2}\right) \right] 
\nonumber \\
&=&\prod_{l=1}^{n}\frac{2}{\pi }\exp \left( -2\left| a_{l}\right|
^{2}\right) .  \label{7h}
\end{eqnarray}
Eq.$\left( \ref{Wvac}\right) $ is the Wigner function of the ``vacuum''
state of the field. The corresponding normalized function in terms of the
field variables $\left\{ y_{l},q_{l}\right\} $ is 
\begin{equation}
W_{vac}\left( \left\{ y_{l},q_{l}\right\} \right) =\prod_{l=1}^{n}\frac{%
\omega _{l}}{\pi %TCIMACRO{\UNICODE[m]{0x127}}
%BeginExpansion
\rlap{\protect\rule[1.1ex]{.325em}{.1ex}}h%
%EndExpansion
}\exp \left( -\frac{\omega _{l}^{2}y_{l}^{2}+q_{l}^{2}}{%
%TCIMACRO{\UNICODE[m]{0x127}}
%BeginExpansion
\rlap{\protect\rule[1.1ex]{.325em}{.1ex}}h%
%EndExpansion
\omega _{l}}\right) .  \label{7s}
\end{equation}

Vacuum expectations of the standard quantum (Hilbert space) treatment become
integrals weighted by the function eq.$\left( \ref{7s}\right) $ in WW. In
particular the expectation value of the Hamiltonian $H_{rad}$ eq.$\left( \ref
{rad}\right) $ reproduces the spectrum eq.$\left( \ref{roZPF}\right) .$

The quantum equations eqs.$\left( \ref{7h}\right) $ and $\left( \ref{rad}%
\right) $ for the ``vacuum'' radiation are identical to the SED eqs.$\left( 
\ref{Wvac}\right) $ and $\left( \ref{ESED}\right) $ for the assumed
radiation filling space. However there is a conceptual difference: In SED eq.%
$\left( \ref{Wvac}\right) $ is interpreted as the probability distribution
of a stochastic field, but in quantum physics it is the Wigner fuction of
the vacuum state.

\subsection{The evolution in phase space}

The (quatum) evolution of any Wigner function $Z\left( \left\{
y_{l},q_{l}\right\} ,t\right) $ of free radiation should be governed by
Moyal eq.$\left( \ref{dW}\right) ,$ but Moyal bracket becomes Poisson
bracket in this case because all derivatives of $H_{rad},$ eq.$\left( \ref
{yq}\right) ,$ of order higher than 2 are nil. Then the evolution is
provided by the following Liouville equation 
\begin{equation}
\frac{\partial Z}{\partial t}=\frac{\partial Z}{\partial y_{l}}\frac{%
\partial H_{rad}}{\partial q_{l}}-\frac{\partial Z}{\partial q_{l}}\frac{%
\partial H_{rad}}{\partial y_{l}}\equiv \left\{ Z,H_{rad}\right\} _{P,rad},
\label{s6}
\end{equation}
where $\left\{ Z,H_{rad}\right\} _{P,rad}$ is the radiation Poisson bracket$%
. $

Now we are in a position to get the Hamiltonian of a system of charged
particles coupled to electromagnetic radiation in the WW formalism. The
Hamiltonian may be derived from the standard quantum one (involving
operators in a Hilbert space) via the transform eq.$\left( \ref{h2}\right) $ 
\cite{FOOP}. It consists of 3 terms, that is 
\begin{equation}
H=H_{part}+H_{rad}+H_{int}.  \label{H}
\end{equation}
A particular instance of mechanical (particles) Hamiltonian $H_{part}\left(
\left\{ \mathbf{x}_{j},\mathbf{p}_{j}\right\} \right) $ was given in eq.$%
\left( \ref{c2}\right) $. The radiation Hamiltonian appears in eq.$\left( 
\ref{rad}\right) $ and the interaction Hamiltonian may be written in terms
of the vector potential, which may be got from eq.$\left( \ref{Hint}\right) $
via theWeyl transform. That is 
\begin{equation}
H_{int}=\sum_{j}[-\frac{e}{mc}\mathbf{p}_{j}\cdot \mathbf{A}\left( \mathbf{x}%
_{j},t\right) +\frac{e^{2}}{2mc^{2}}\mathbf{A}\left( \mathbf{x}_{j},t\right)
\cdot \mathbf{A}\left( \mathbf{x}_{j},t\right) ],  \label{5}
\end{equation}
where the sum runs over the different particles, $\mathbf{x}_{j}$ and $%
\mathbf{p}_{j}$ being the position and momentum of particle \textit{j} at
time t. It is straightforward to write the vector potential in terms of the
field variables, but it is not necessary for our purpose and I skip it.

After that we may treat the evolution of the system of particles plus
radiation in a unified way analogous to the treatment of the particles
alone. That is we define $F\left( \left\{ \mathbf{x}_{j},\mathbf{p}%
_{j},y_{l},q_{l}\right\} ,t\right) $ to be the joint Wigner function in the
space consisting of the phase space of the particles times the space spanned
by the numerable set of field coordinates $\left\{ y_{l},q_{l}\right\} .$
The evolution is given by the following generalized Moyal equation 
\begin{equation}
\frac{\partial F}{\partial t}=\left\{ F,H_{part}+H_{int}\right\}
_{M}+\left\{ F,H_{rad}\right\} _{P},  \label{7}
\end{equation}
where the Moyal bracket (subindex M) reduces to Poisson's (subindex P) in
the second term because $H_{rad}$ is quadratic in the field amplitudes. Eq.$%
\left( \ref{7}\right) $ might be the starting point for calculations in
nonrelativistic QED treated in the WW formalism.

\section{The approximation leading to analogy with SED}

\subsection{The Liouville equation}

Solutions of the quantum evolution equation of nonrelativistic QED in the WW
formalism, eq.$\left( \ref{7}\right) ,$ will not be studied in the present
article. We shall deal only with the approximation to first order in Planck
constant. It involves neglecting all terms of the expansion defining the
Moyal bracket in eq.$\left( \ref{7}\right) ,$ except the first one. Actually
taking into account that the Moyal expansion consists of terms with even
powers of $%TCIMACRO{\UNICODE[m]{0x127}}
%BeginExpansion
\rlap{\protect\rule[1.1ex]{.325em}{.1ex}}h%
%EndExpansion
$, the approximation means neglecting terms of order O$\left( 
%TCIMACRO{\UNICODE[m]{0x127}}
%BeginExpansion
\rlap{\protect\rule[1.1ex]{.325em}{.1ex}}h%
%EndExpansion
^{2}\right) $. The radiation and the interaction Hamiltonians, eqs.$\left( 
\ref{yq}\right) $ and $\left( \ref{5}\right) $ respectively, do not contain $
%TCIMACRO{\UNICODE[m]{0x127}}
%BeginExpansion
\rlap{\protect\rule[1.1ex]{.325em}{.1ex}}h%
%EndExpansion
,$ but Planck constant appears in the vacuum state eq.$\left( \ref{7s}%
\right) $ whence the approximation of eq.$\left( \ref{7}\right) $ just
described, leading to 
\begin{equation}
\frac{\partial F}{\partial t}=\left\{ F,H_{part}+H_{int}\right\}
_{P}+\left\{ F,H_{rad}\right\} _{P}=\left\{ F,H\right\} _{P},
\label{18}
\end{equation}
may be labelled an approximation to first order in $
%TCIMACRO{\UNICODE[m]{0x127}}
%BeginExpansion
\rlap{\protect\rule[1.1ex]{.325em}{.1ex}}h%
%EndExpansion
.$ It is formally a classical Liouville equation for particles and field
toghether.

The relevant point is that eq.$\left( \ref{18}\right) $ is equally valid for
both classical electrodynamics (CED) and ``nonrelativistic QED in the WW
formalism approximated to order $O\left( %TCIMACRO{\UNICODE[m]{0x127}}
%BeginExpansion
\rlap{\protect\rule[1.1ex]{.325em}{.1ex}}h%
%EndExpansion
\right) "$. For simplicity, and also clarity, the expression in inverted
commas will be labelled QEDWC in the following, where C stands for the
approximation leading from eq.$\left( \ref{7}\right) $ to the \textit{%
classical} Liouville eq.$\left( \ref{18}\right) .$ Thus the evolution
equations are classical in both CED and QEDWC, that is both rest on
Maxwell-Lorentz theory. However there is a difference in the initial
conditions appropriate for the integration of the (Liouville) partial
differencial eq.$\left( \ref{18}\right) .$ In CED the initial state
corresponds to absence of radiation, but in QEDWC it should be the quantum
vacuum state for the radiation, that is ZPF given by eq.$\left( \ref{7s}%
\right) $. For the particles the initial state may be any probability
distribution in phase space in CED, but it should be a Wigner function in
QEDWC. I stress that the Wigner function for particles is \textit{not}
interpreted as a probability distribution too.

QEDWC is a mixture of classical evolution with quantum initial conditions.
The interest for us is that SED has formal similarity with QEDWC. In fact eq.%
$\left( \ref{18}\right) $ is valid in both cases and the initial state for
the radiation should be also the same, that is the ZPF eq.$\left( \ref{7s}%
\right) ,$ but maybe not for the particles. Also there is a conceptual
difference between SED and QEDWC. In SED eq.$\left( \ref{7s}\right) $ 
\textit{is interpreted} as a real random radiation filling space whence eq.$%
\left( \ref{18}\right) $ is the evolution equation of particles interacting
with the random field ZPF. Hence we may state

\begin{proposition}
From a formal (mathematical) point of view SED is similar to QEDWC except
that the initial state of the particles may be any probability distribution
in phase space in the former but it should be a Wigner function in the
latter.
\end{proposition}

This justifies the success of SED when dealing with linear systems, that is
particle Hamiltonians quadratic at most in the coordinates and momenta of
the particles. In fact, in this case no approximation takes place when we go
from eq.$\left( \ref{7}\right) $ to eq.$\left( \ref{18}\right) $ because
only the first term contributes in the Moyal eq.$\left( \ref{dW}\right) $.
In sharp contrast we should expect that SED fails badly for Hamiltonians not
quadratic. Indeed the neglect of terms O$\left( 
%TCIMACRO{\UNICODE[m]{0x127}}
%BeginExpansion
\rlap{\protect\rule[1.1ex]{.325em}{.1ex}}h%
%EndExpansion
^{2}\right) ,$ in going from Moyal to Poisson brackets, may produce errors
of order $%TCIMACRO{\UNICODE[m]{0x127}}
%BeginExpansion
\rlap{\protect\rule[1.1ex]{.325em}{.1ex}}h%
%EndExpansion
^{2},$ more properly order $(%TCIMACRO{\UNICODE[m]{0x127}}
%BeginExpansion
\rlap{\protect\rule[1.1ex]{.325em}{.1ex}}h%
%EndExpansion
\omega /E)^{2}$, where the ratio of energy $E$ by frequency $\omega $ may be
taken as the typical action. In the microscopic (quantum) domain $
%TCIMACRO{\UNICODE[m]{0x127}}
%BeginExpansion
\rlap{\protect\rule[1.1ex]{.325em}{.1ex}}h%
%EndExpansion
\omega /E$ is of order unity, whence we conclude that SED applied to
nonlinear systems is a rather bad approximation to quantum theory, which has
been shown in actual calculations \cite{dice}, \cite{book}, \cite{arxiv}.

\subsection{The equations of motion in QED and in SED}

In SED it is common to start from eq.$\left( \ref{ode}\right) $ rather than
eq.$\left( \ref{18}\right) .$ In the following I study the relation between
these two approaches. This will prove more clearly the formal analogy
between QEDWC and SED.

In classical dynamics there is a close connection between the Liouville
equation and the Hamilton equations of motion. The latter are coupled
ordinary differential equations that, when integrated, provide the final
positions and momenta of the particles in terms of the initial ones. The
solutions may be written 
\begin{equation}
\mathbf{x}_{j}\left( t\right) =\mathbf{X}_{j}\left( \left\{ \mathbf{x}%
_{k}\left( 0\right) ,\mathbf{p}_{k}\left( 0\right) \right\} ,t\right) ,%
\mathbf{p}_{j}\left( t\right) =\mathbf{P}_{j}\left( \left\{ \mathbf{x}%
_{k}\left( 0\right) ,\mathbf{p}_{k}\left( 0\right) \right\} ,t\right) ,
\label{30}
\end{equation}
where $\left\{ \mathbf{X}_{j},\mathbf{P}_{j}\right\} $ are functions
obtained from the integration of the Hamilton equations. In principle from
eqs.$\left( \ref{30}\right) $ we might get the evolution of a distribution
in phase space, which would amount to solving the Liouville equation. In
fact, if we are given a function $F_{t}$ in phase space at time $t$ we may
get the function at time $0$ as follows: 
\begin{eqnarray}
F_{t}\left( \left\{ \mathbf{x}_{j}\left( t\right) ,\mathbf{p}_{j}\left(
t\right) \right\} \right)  &=&F_{t}\left( \mathbf{X}_{j}\left( \left\{ 
\mathbf{x}_{k}\left( 0\right) ,\mathbf{p}_{k}\left( 0\right) \right\}
\right) ,\mathbf{P}_{j}\left( \left\{ \mathbf{x}_{k}\left( 0\right) ,\mathbf{%
p}_{k}\left( 0\right) \right\} \right) \right)   \nonumber \\
&=&F_{o}\left( \left\{ \mathbf{x}_{j}\left( 0\right) ,\mathbf{p}_{j}\left(
0\right) \right\} \right) .  \label{35}
\end{eqnarray}
Similarly we might obtain $F_{t}$ from $F_{o}$ using the equations obtained
inverting eqs.$\left( \ref{30}\right) .$

An analogous relation exists bewteen the Liouville and the Hamilton
equations in the two theories that we are considering in this article,
namely CED and QEDWC. In fact they are formally classical dynamical problems
for the variables of both the particles and the electromagnetic field. In
the following I will compare both theories via the Hamilton equations of
motion.

Let us start studying the solution of dynamical equations for charged
particles in classical electrodynamics (CED). From the Hamiltonian eq.$%
\left( \ref{H}\right) $ we have derived eq.$\left( \ref{18}\right) $ and it
is also easy to derive the canonical equations of motion. For simplicity I
will study a single particle with Hamiltonian 
\[
H_{part}=\left[ \frac{\mathbf{p}^{2}}{2m}+V\left( \mathbf{x}\right) \right]
. 
\]
For the particle coupled to radiation we obtain from eq.$\left( \ref{H}%
\right) $ the following Hamilton equations 
\begin{eqnarray}
\frac{d}{dt}\mathbf{x} &=&\frac{\mathbf{p}}{m}+\frac{dH_{int}}{d\mathbf{p}},%
\frac{d}{dt}\mathbf{p}=-\frac{d}{d\mathbf{x}}V\left( \mathbf{x}\right) , 
\nonumber \\
\frac{d}{dt}y_{l} &=&q_{l}+\frac{dH_{int}}{dq_{l}},\frac{d}{dt}q_{l}=-\omega
_{l}^{2}y_{l}-\frac{dH_{int}}{dy_{l}}.  \label{CED}
\end{eqnarray}
where $H_{int}$ was defined in eq.$\left( \ref{5}\right) .$ From the former
two equations we get 
\begin{equation}
m\frac{d^{2}\mathbf{x}}{dt^{2}}\mathbf{=-}\frac{dV\left( \mathbf{x}\right) }{%
d\mathbf{x}}+e\mathbf{E},  \label{CED3}
\end{equation}
where $\mathbf{E}$ is the electric field of the radiation acting on the
particle. The action of the magnetic field has been neglected, a usual
approximation in nonrelativistic theory

Classical electrodynamics corresponds to solving eqs.$\left( \ref{CED}%
\right) $ with nil radiation at the initial time, that is 
\[
y_{l}^{0}=q_{l}^{0}=0\text{ for all }l.
\]
Solving the coupled differential eqs.$\left( \ref{CED}\right) $ is lengthy,
but the subject has been studied from long ago and the result is well known.
In fact with a fair approximation one obtains \cite{Milonni} 
\begin{equation}
m\frac{d^{2}\mathbf{x}}{dt^{2}}\mathbf{=-}\frac{dV\left( \mathbf{x}\right) }{%
d\mathbf{x}}+e\mathbf{E}_{RR},\mathbf{E}_{RR}=\frac{2e}{3c^{3}}\stackrel{...%
}{\mathbf{x}},  \label{CED1}
\end{equation}
where $\mathbf{E}_{RR}$ is named radiation reaction field and $e\mathbf{E}%
_{RR}$ it is interpreted as the effective friction force acting on the
particle due to the radiation emitted by it in its accelerated motion. (I
point out that the approximation involving $\stackrel{...}{\mathbf{x}}$
presents well known problems like the existence of runaway unphysical
solutions, that I will not discuss here). Incidentally the radiation
reaction causes that matter is unstable according to classical
electrodynamics. The inestability does not appear in QED as is well known
and it is instructive to see how this happens in QEDWC, which is shown in
the following.

As explained above \textit{the same Hamiltonian} may be used in both QEDWC
and CED. The difference does not pertain to the equations of motion, that
are formally identical, but to the required initial conditions. QEDWC is a
quantum theory whence the initial condition should be a Wigner function
representing the state of both particles and field. In analogy with
classical dynamics the evolution of the Liouville function might be obtained
in principle via the solution of the Hamilton equations of motion, similar
to the passage from eq.$\left( \ref{30}\right) $ to eq.$\left( \ref{35}%
\right) .$ We should start solving eqs.$\left( \ref{CED}\right) $ in order
to get the variables at time $t$, that is 
\begin{equation}
\mathbf{x}\left( \lambda ,t\right) ,\mathbf{p}\left( \lambda ,t\right)
,y_{l}\left( \lambda ,t\right) ,q_{l}\left( \lambda ,t\right) ;\lambda
\equiv (\mathbf{x}_{0},\mathbf{p}_{0},\left\{ y_{l}^{0},q_{l}^{0}\right\} )
\label{lambda}
\end{equation}
$\lambda $ being the set of initial conditions. Hence we might obtain in
principle the evolution of the Wigner function of both particles and field, F%
$\left( \mathbf{x},\mathbf{p},\left\{ y_{l},q_{l}\right\} ,t\right) $ eq.$%
\left( \ref{18}\right) ,$ from the initial Wigner function F$\left( \lambda
,0\right) .$

The problem of QEDWC is to find the electric field $\mathbf{E}\left( \lambda
,t\right) $ that results from the solution of eqs.$\left( \ref{CED}\right) $
with the initial conditions given in eqs.$\left( \ref{lambda}\right) .$ The
solution is lengthy but a plausible approximation is that the initial
conditions $\left\{ y_{l}^{0},q_{l}^{0}\right\} $ for the field just gives
rise to an electric field $\mathbf{E}_{ZPF}$ additional to $\mathbf{E}_{RR}$
(see eq.$\left( \ref{CED1}\right) )$. That is the electric field of eq.$%
\left( \ref{CED3}\right) $ should lead to 
\begin{equation}
m\stackrel{..}{\mathbf{x}}=-\frac{dV\left( \mathbf{x}\right) }{d\mathbf{x}}+%
\frac{2e^{2}}{3c^{3}}\stackrel{...}{\mathbf{x}}+e\mathbf{E}_{ZPF}\left(
\lambda ,t\right) .  \label{SED1}
\end{equation}
Furthermore we may approximate the evolution of $\mathbf{E}_{ZPF}\left(
\lambda ,t\right) $ by a \textit{free evolution} of the field with initial
conditions $\left\{ y_{l}^{0},q_{l}^{0}\right\} .$

If we solved eq.$\left( \ref{SED1}\right) $ for a family of initials
conditions $\lambda \in \Lambda $ we would get a family of solutions 
\begin{equation}
\mathbf{x=x}\left( \lambda ,t\right) ,\mathbf{p=p}\left( \lambda ,t\right) ,
\label{SED2}
\end{equation}
whence in principle we might obtain the Wigner function of both particle and
radiation at time $t$ from the initial Wigner function. That initial Wigner
function should consist of the product of a Wigner fuction for the particle,
that may be chosen at will, times the Wigner function of the radiation which
should be the vacuum Wigner function eq.$\left( \ref{7s}\right) .$ This
would provide the solution of the problem in the QEDWC formalism in
principle.

The functions\textbf{\ }that appear in eq.$\left( \ref{SED2}\right) $ have
formal analogy with stochastic processes, which are currently defined as
functions of time $t$ and ``chance'' $\lambda $. This suggests that we could
solve more efficiently the problem assuming that $\lambda $ is actually a
multidimesional random variable with probability distribution eq.$\left( \ref
{7s}\right) $ for the field. In QEDWC the initial condition for the particle
should be a Wigner function, but we may consider a similar theory where we
use definite initial conditions for the particle, say $\mathbf{x}\left(
0\right) \mathbf{=x}^{0},\mathbf{p}\left( 0\right) \mathbf{=p}^{0}$. This is
precisely the choice made in SED, where eqs.$\left( \ref{SED2}\right) $ are
taken as stochastic processes and eq.$\left( \ref{SED1}\right) $ leads to
the stochastic differential eq.$\left( \ref{ode}\right) .$

\section{The problems of interpretation}

Quantum mechanics works yet despite its unprecedented success there is
notorious disagreement for the interpretation. Everybody who has learned
quantum mechanics agrees how to use it but we do not understand the meaning
of the strange conceptual apparatus that each of us uses so effectively to
deal with our world. For decades some physicists have been searching for
modifications of quantum mechanics that either maintain its testable
predictions or lead to changes too small to have yet been observed. Such
modifications are motivated not by failures of the existing theory, but by
philosophical discomfort with one or another of the prevailing
interpretations of that theory. One of these attempts has been SED.

In my view a correct understanding of quantum theory should acknowledge that
``the physical concepts with which the theory operates are intended to
correspond with the objective reality, and by means of these concepts we
picture this reality to ourselves'' \cite{EPR}, an epistemology which I
would label \textit{realistic interpretation}. SED was, from the very
beginning in the 1960', an attempt to get a realistic interpretation of
quantum mechanics or even to replace it by a deeper theory. However the
difficulties for a generalization of SED beyond nonrelativistic
electrodynamics and the failure to interpret even simple systems, like the
hydrogen atom\cite{Neu}, led many authors to less ambitious expectations. In
my present opinion SED is a kind of approximation to nonrelativistic QED
with validity in a narrow domain.

This does not mean that the effort to develop SED has been useless. At
present I think that it has been a clue or guide in the search for a better
understanding of quantum theory, which now I see as a stochastic theory
resting on the interpretation of the quantum vacuum fields like real
stochastic fields\cite{FOOP}. Furthermore the interpretation of the
electromagnetic field within the Weyl-Wigner formalism provides a realistic
picture for many quantum effects\cite{Foundations}.

\section{Data Availability Statement:}

No Data associated in the manuscript.

\end{document}